\documentclass{aa}
\usepackage[pdftex,unicode]{hyperref}
\usepackage{graphicx,natbib,amssymb}
\bibpunct{(}{)}{;}{a}{}{,}

\def\1p5{1.5DF}

\def\flux{\rm erg~s$^{-1}$~cm$^{-2}$}
\def\lum{\rm erg/s}

\begin{document}

\title{On the nature of the break in the X-ray luminosity function of
  low-mass X-ray binaries.}

\author{M.~Revnivtsev\inst{1,2}, K.~Postnov\inst{3},
  A.~Kuranov\inst{3}, H.~Ritter\inst{4}}

\institute{Excellence Cluster Universe, Technische Universit\"at
  M\"unchen, Boltzmannstr.2, 85748 Garching, Germany
\and Space Research Institute, Russian Academy of Sciences,
Profsoyuznaya 84/32, 117997 Moscow, Russia
\and Sternberg Astronomical Institute, Moscow State University,
Universitetskij pr., 13, 119992, Moscow, Russia
\and Max-Planck-Institute f\"ur Astrophysik, Karl-Schwarzschild-Str.
1, D-85740 Garching bei M\"unchen, Germany
}
\date{}
\authorrunning{Revnivtsev et al.}
\titlerunning{On the nature of break in the X-ray luminosity function
  of LMXBs} 

\abstract{We analyze a flux-limited sample of persistent and bright
  (with 2-10 keV fluxes exceeding $1.4\times10^{-10}$ \flux\ )
  low-mass X-ray binaries (LMXBs) in our Galaxy. It is demonstrated
  that the majority of binary systems with X-ray luminosities below
  $\log L\textrm{(erg/sec)}\sim 37.3$ have unevolved secondary
  companions (except for those with white dwarf donors), while systems
  with higher X-ray luminosity predominantly harbor giant donors. Mass
  transfer in binary systems with giants significantly shortens their
  life time thus steepening the X-ray luminosity function of LMXBs at
  high luminosity. We argue that this is the reason why the LMXB
  luminosity function constructed in the last years from observations
  of sources in our and distant galaxies demonstrates a break at
  $\log L\textrm{(erg/sec)}\sim 37.3$.}
\maketitle

\section{Introduction}

Binary systems with low-mass secondary companions to a compact star
(black hole or neutron star) were discovered in the 1960s at the dawn
of X-ray astronomy. All-sky surveys performed by different orbital
X-ray observatories  (UHURU, HEAO1, Ariel V, etc.) provided us with a
relatively large sample of such objects in our Galaxy. The advent of
focusing X-ray telescopes with an angular resolution of arcseconds
initiated studies of such low-mass X-ray binaries (LMXBs), first in
nearby galaxies, like M31 \citep{trinchieri91,primini93}, and then in
more distant galaxies \cite[see e.g. reviews in ][]{kim04,gilfanov04}.

The accretion luminosity $L_x$ of a persistent (sub-Eddington) LMXB is directly
proportional to the mass transfer rate $\dot M$ from the secondary
star. According to the standard theory of binary star evolution, the
persistent mass transfer via the inner Lagrangian point in a close
binary is due to an increase of the size of the donor star relative to
its Roche lobe. In an LMXB, this can be done either by decreasing the
size of the Roche lobe due to loss of the orbital angular momentum
\cite[see e.g.][]{paczynski81,verbunt81}, or by increasing the radius
of the star as a consequence of its nuclear evolution
\citep{webbink83,taam83,ritter99}. The observed properties of
different populations of LMXBs can be used to test the  models of
binary evolution.

The X-ray luminosity function (LF) is an important characteristic of
the LMXB population. In other galaxies it is directly constructed from
measured X-ray fluxes down to a luminosity $L_{\rm  x}\sim 10^{37}$
\lum\ and can be fitted by a power law  $dN/d \log L\propto
L^{-0.8...-1.2}$  with a steepening at luminosities exceeding the
Eddington limit for accreting neutron stars ($\log L_{\rm x}>38.5$)
\citep{kim04}. However, recently it became possible to construct the
luminosity functions of LMXBs in nearby galaxies, like M31 or Cen A,
down to much smaller luminosities of the order of $10^{36}$ \lum\ 
\citep{primini93,gilfanov04,voss06,voss07,voss09}. Over such a wide
luminosity interval the luminosity function of LMXBs can no longer be
described by a single power law and demonstrates a characteristic
break at $\log L <37.3$ \citep{gilfanov04}. A similar result down to
luminosities $L_{\rm x}\sim 10^{35}$ erg/sec was obtained for galactic
LMXBs from all-sky surveys \citep{grimm02,gilfanov04} and the survey
of the Galactic bulge \citep{revnivtsev08}. The statistical
significance of the LF break in the above-mentioned works is high,
therefore the existence of the break requires physical explaination. 
One of the clearest case of the flattening of the LMXB LF at low 
luminosities can be seen in the work of \cite{revnivtsev08}, where the 
LMXB candidates in the bulge of our Galaxy were traced down to 
luminosities $10^{35}$ erg/sec, almost unreachable for LMXBs in outer
galaxies. 

Different explanations have been proposed for the origin of the break
in the observed LF of LMXBs. For example, the break in the LF might be
caused by the change of the dominant orbital angular momentum loss
mechanism from magnetic stellar wind \citep{verbunt81} to
gravitational wave emission \cite[see e.g.][]{paczynski81} in the
population of LMXBs with low-mass main sequence secondaries
\citep{postnov05}. It was also noted already quite long ago that the
most luminous LMXBs with high mass transfer rates ($\dot{M}>10^{-8}$
$M_\odot$/year), could harbor giant donors \cite[see
  e.g.][]{webbink83}, and this may underly the difference between the
low- and high- luminosity sources. The study of the LMXB LF in
galaxies using population synthesis methods could potentially be quite
powerful \cite[see e.g.][]{fragos08, kim09}. However, this approach
involves a number of uncertain parameters of binary star evolution, so
it is hard to make firm conclusions based on the population synthesis
simulations only.

In our paper we make an attempt to identify the main physical reason
for the origin of the observed break of the broad band luminosity
function of LMXBs.

We show that the break in the LF of LMXBs is probably caused by 
different types of donor stars: evolved secondaries (giants) at 
luminosities above the break at  $\log L\textrm{(erg/sec)}\sim 37.3$, 
and main sequence stars at lower luminosities. The correctness of our 
conclusion might be checked by direct calculations of the LMXB 
properties using the methods of population synthesis, which we plan 
to do in the future.

\section{A Flux limited sample of persistent LMXBs in the Galaxy}

We consider only persistent LMXBs. This selection can be relatively
easily done for galactic LMXBs, but in general this can be a difficult
task for LMXBs in outer galaxies, since they are typically observed
only during short time intervals. Monitoring of X-ray sources in our
Galaxy shows that at any particular time there are only a few bright
transients \cite[see e.g.][]{remillard09}, whose effect on the
instantaneous LF of all galactic LMXBs is rather small.

Moreover, the properties of the observed luminosity distribution of
individual variable LMXBs were found to not affect the shape of the
instantaneous LF of the galactic sample \citep{postnov05}. Therefore
the results of our present study can also be applied to the
instantaneous sample of LMXBs in distant galaxies.

We study a representative sample of persistent galactic LMXBs with
X-ray luminosities $L_{\rm x}>10^{36}$ \lum\  and compare their
properties above and below the break. Our goal does not require the 
sample to be volume limited or complete, because we are not interested 
in statistics (which was extensively studied in works of other
authors), but rather we look for a correspondence of the time averaged
X-ray luminosity of sources to some parameters of their binaries.

To select persistent Galactic sources we can analyze any existing
X-ray survey. However, such a survey should fulfill the following 
requirements: 1) it should cover as much as possible of the Galactic 
plane region, where the majority of the LMXBs are located, 2) it
should have an angular resolution of better than $\sim2-3^\circ$ in
order to avoid confusion within the Galactic plane, 3) it should be 
performed in the energy band $\sim1-20$ keV where Galactic LMXBs emit 
most of their bolometric luminosity. These requirements significantly 
shorten the list of usable surveys. In particular, the sky survey of 
the RXTE observatory \citep{revnivtsev04} does not cover the region 
of the Galactic plane, and the scans of the RXTE observatory over the 
Galactic bulge and the Galactic plane \citep{markwardt00} have not 
(yet) been used to perform a sky survey, rather it is used to measure 
the fluxes of preselected list of sources. Surveys carried out by the 
ASCA \citep{sugizaki01,sakano02} and BeppoSAX \citep{sidoli01} 
observatories cover only a small fraction of the Galaxy. Even smaller 
sky areas are covered by  the {\it Chandra} and XMM observatories. The 
survey of the INTEGRAL observatory \citep{krivonos07} covers the whole 
Galaxy, but it is done at hard X-ray energies, and is not fully
suitable for our purposes (however, we will use it for additional
checks of "persistency" of selected sources).

So the persistent X-ray sources detected by UHURU \citep{forman78}
turn out to be most suitable for our study; as an additional check of 
their persistent behavior we have also examined their presence in the 
INTEGRAL all sky survey  \citep{krivonos07}. We selected only sources 
with 2-10 keV fluxes above $1.4\times 10^{-10}$ \flux\, which ensures 
that their luminosity is not lower than $2-3\times 10^{36}$ \lum\ up
to distances of 12 kpc (i.e. up to the further edge of the Galactic 
bulge). This allows us to conclude that we do not miss brighter
sources at least in more than a half of the Galaxy. We also included 
in our sample the black-hole binary GRS 1915+105, which was not seen 
by UHURU during its operation time (1971-1973), but now remains 
persistently bright since its appearance in 1992 \citep{castro92}, and 
excluded the source GX 1+4, which is known to be accreting via a 
stellar wind \citep{hinkle06} and not via Roche lobe overflow which we 
consider here. The fluxes assigned to GRS~1915+105 and GS~1826-24 were 
taken from measurements of the All Sky Monitor of the RXTE observatory 
averaged over 1996-2009 (see. e.g. 
{\tt http://xte.mit.edu/asmlc/ASM.html} ).

Energy fluxes from the selected sources were calculated from observed 
UHURU count rates, assuming that the Crab nebula count rate  in the 
energy band 2--10 keV corresponds to the energy flux
$2.22\times10^{-8}$ \flux\ .

In Table 1 we present the list of sources with values of their orbital 
periods and estimates of their distances and luminosities.

\begin{table}[htb]
\caption{The brightest persistent low mass X-ray binaries in the
  Galaxy with fluxes $>1.4\times10^{-10}$ \flux\ measured by UHURU
\citep{forman78}. Orbital periods of systems are adopted from
\cite{ritter03} and \cite{liu07}.}
\small
\begin{center}
\begin{tabular}{l|c|c|c}
Name     &  $L_{\rm x}$ (2-10 keV)&Dist, kpc&Period,h\\
         &   $10^{37}$ erg/sec&\\
\hline
Sco X-1  &       37.4 &             2.8$^1$ &            18.94\\
GX 5-1   &       26.1 &            9$^2$ &           \\
GRS 1915+105&    26.0 &           11$^3$&          739.20\\
Cyg X-2  &       19.4 &        11.6$^5$ &           236.27\\
GX340+0  &       18.2 &         12$^4$ &           \\
GX 349+2 &       15.3 &       8.5$^6$&            22.5\\
GX 17+2  &       14.9 &      7.5$^{7}$ &             \\
GX 9+1   &       8.7 &              7.2$^{7}$ &         \\
4U1820-30&       5.7 &               8$^{8}$&           0.19\\
Ser X-1  &       5.5 &             8.4$^{9}$ &       12.96\\
GX 13+1  &       5.5 &            6.9$^{10}$  &     601.7\\
4U1735-44&       4.9 &             9.1$^{11}$ &       4.65\\
GX 3+1   &       3.4 &              4.5$^{12}$&       \\
4U1624-49 &      3.1 &             15$^{13}$ &          20.9\\
4U1636-53&       2.4 &            5.9$^{10}$ &       3.79\\
GX 9+9   &       1.8 &              5.0$^{7}$&       4.20\\
 4U1746-37 &       1.4 &          11.0$^{8}$&          5.16\\
 4U1705-32 &      1.18 &       13$^{14}$&               \\
 1A1742-294 &      0.81 &            8.5$^{15}$&               \\
 4U1254-69 &       0.79 &          11$^{16}$ &            3.93\\
 4U0513-40 &       0.74 &            12.1$^{8}$&         0.28\\
 4U1823-00 &       0.61 &          6.3$^{17}$ &             3.19\\
 4U1915-05 &       0.43 &         8.8$^{8}$&             0.83\\
 4U0614+09 &       0.34 &              3.2$^{18}$&              0.81\\
 4U1702-42 &       0.32 &               6.2$^{19}$ &                  \\
 4U1626-67 &       0.32 &            8$^{20}$ &            0.69\\
 GS1826-24 &       0.31 &            6.0$^{21}$ &             2.25$^{22}$\\
 4U1708-40 &       0.27 &          8$^{15}$ &               \\
 4U1543-62 &       0.26 &             7.0$^{8}$&             0.30\\
 4U1724-30 &       0.17 &               9.5$^{8}$ &                   \\
 4U1850-08 &       0.17 &             8.2$^{8}$&             0.34\\
 4U1812-12 &       0.09 &        4.0$^{19}$ &                \\
 4U1556-60 &       0.09 &           4.0$^7$&             \\
 4U1822-37 &       0.04$^{*}$&        2.5$^{23}$ &        5.57\\
\end{tabular}
\end{center}
\begin{list}{}
\scriptsize
\item (1) -- \citealt{bradshaw99},(2) -- \citealt{jonker00}, (3) -- \citealt{fender99}, (4) -- \citealt{vanparadijs95}, (5) -- \citealt{smale98},  (6) -- \citealt{wachter96}, (7) -- \citealt{christian97},  (8) -- \citealt{kuulkers03}, (9) -- \citealt{ebisuzaki84}, (10) -- \citealt{band99}, (11) -- \citealt{augusteijn98},  (12) -- \citealt{kuulkers00}, (13) -- \citealt{xiang07}, (14) -- \cite{intzand05}, (15) - adopting distance to the Galactic Center, (16) -- \cite{courvoisier86}, (17) -- \citealt{shahbaz07}, (18) -- \citealt{kuulkers10}, (19) -- \citealt{jonker04}, (20) -- \citealt{chakrabarty98}, (21) -- \citealt{heger07}, (22) -- Mescheryakov et al. 2010, (23)-- \citealt{mason82}
\item $^*$ -- the source belongs to a group of so called accretion disk corona sources, in which we do not see the direct emission of the innermost parts of the accretion flow, therefore the observed X-ray luminosity of such a source is only a part of its intrinsic X-ray luminosity.
\end{list}
\end{table}

\section{Results}

From Table 1 and Fig.1 it is clearly seen that all systems with X-ray 
luminosities exceeding $\sim5\times 10^{37}$ \lum\, for which we have 
information about their orbital periods, apparently have large sizes. 
This (in combination with the assumption that the companion star fills 
its Roche lobe) means that donor stars in these systems must have
large radii and thus are evolved stars (subgiants or giants). In
several cases this was shown directly from the observed spectra of the 
optical counterparts (Sco X-1, \citealt{band97}; GX 349+2, 
\citealt{band99}; Cyg X-2, \citealt{casares98}; GX 5-1, 
\citealt{band03}; GRS 1915+105, \citealt{greiner01}; Cir X-1, 
\citealt{jonker07}).

\begin{figure}
\includegraphics[width=\columnwidth]{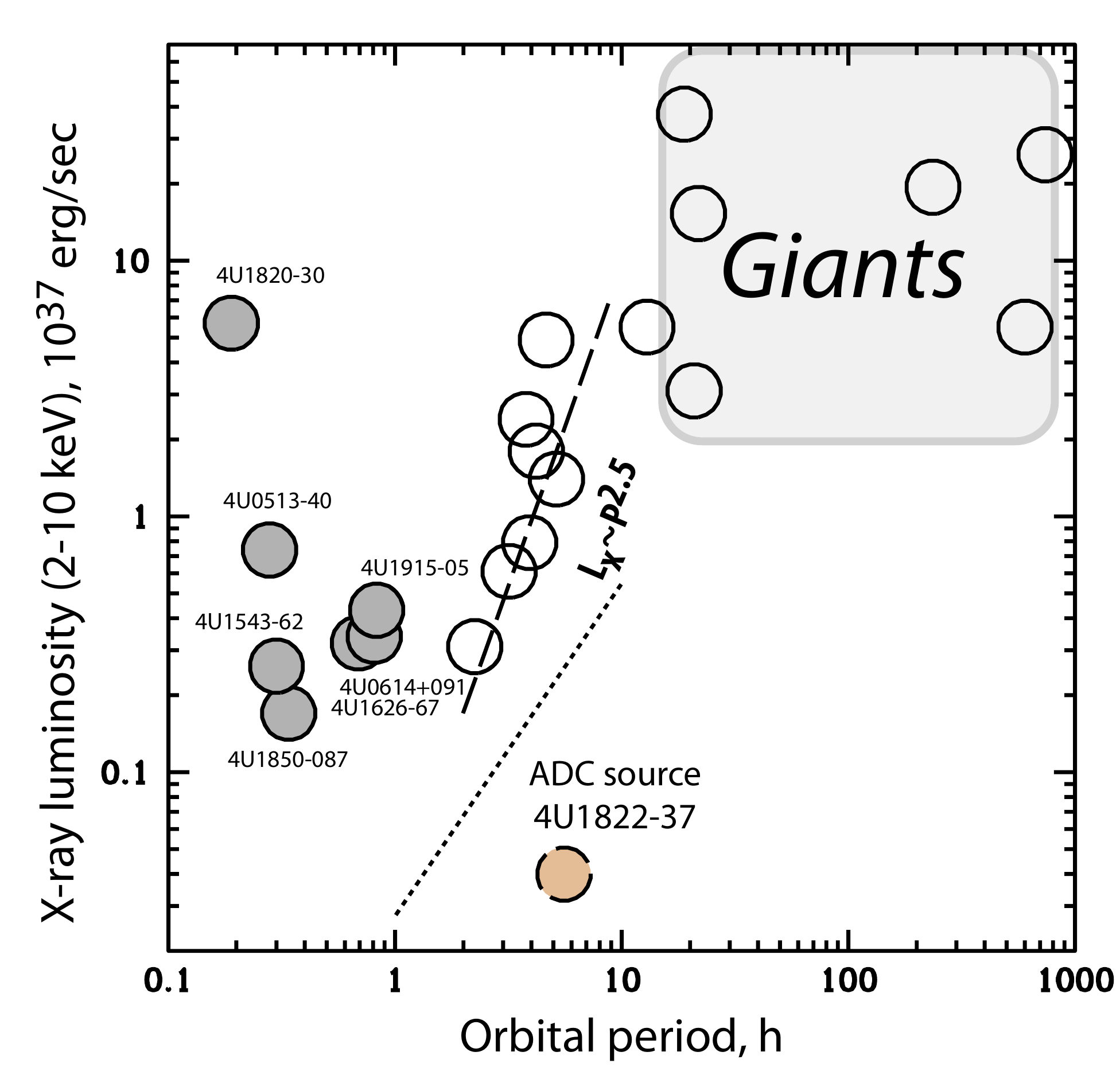}
\caption{Positions of the LMXBs from Table 1 on the orbital period -- 
X-ray luminosity diagram. Filled circles denote the positions of 
peculiar systems: gray circles show the positions of binaries with 
white dwarf or hydrogen-deficit accretors, the dashed circle shows 
the position of 4U1822-37 in which we see only a fraction of the total 
X-ray luminosity due to its nearly edge-on orientation (a so-called 
accretion disk corona source). The region, occupied by binary systems 
with a giant donor is shown by the shaded square. The dashed line 
shows the slope $L_{\rm x}\propto P^{2.5}$, derived by \cite{iben84}
for the mass tranfer rate in binaries in which angular momentum loss 
is driven by a magnetic stellar wind from a main sequence donor. The 
dotted line shows the boundary of persistency according to the 
thermal-viscous disk instability for binaries with low-mass main 
sequence donors $L_x/10^{37}$ erg/sec $\sim0.025~P_{\rm hr}^{1.4}$
from eqn.32 of \cite{dubus99}, where we adopted $M_1=1.4M_\odot$, 
$M_2=0.4M_\odot$ and $L_{\rm x}=0.1\dot{M}c^2$.}
\label{plot}
\end{figure}

To the contrary, there are virtually no persistent LMXBs with X-ray 
luminosities lower than $\sim10^{37.5}$ \lum\ and orbital periods
longer than 5-10 hours \cite[see also][]{ritter03,liu07}. This is 
consistent with the assumption that donor stars in these systems are 
main sequence stars (or degenerate dwarfs, if the orbital period is 
smaller than $\sim1$ hour). For cataclysmic variables with accreting 
white dwarfs and orbital periods less than $\sim$10 hours (which are 
much more numerous in the Galaxy and can be found closer to the Sun, 
so we can study them in much greater details) this statement has been 
solidly supported by observations \citep{smith98}. Therefore, {\sl 
we conclude from observations that at luminosities} 
$L_{\rm x}\sim10^{37.5}$ \lum\ {\sl there is a transition from LMXBs
  with predominantly giant donors to binaries with predominantly
  main-sequence donors}.

\section{A simple model of the LMXB luminosity function}

Numerous studies have been made in the past to describe the evolution 
of the LMXB population using specific models of LMXB evolution. They 
were mostly aimed at understanding the evolution of the total number 
of galactic LMXBs without distinguishing their X-ray luminosities 
\cite[see e.g][]{white98,ghosh01}. Attempts to describe the LMXB LF
can be found in \cite{webbink83}, but this work concentrated only on 
the bright end of the LF and does not consider the effects of magnetic 
braking on the period/luminosity distribution of binaries. The
importance of the magnetic braking for relatively short period systems 
was emphasized, for example, in papers by \cite{pylyser88,cote89}. In 
these papers, an indication of transition from a population of bright 
systems dominated by nuclear evolution of donor stars to a population 
of systems in which mass transfer is sustained by the magnetic braking 
can be found. At that time it was impossible to  make a quantitative 
comparison of the model predictions with the observed samples of LMXBs 
due to the lack of sensitive galactic X-ray surveys. So these studies 
were mostly focused on the exploration of evolutionary tracks leading 
to the formation of particular objects, like rotation-powered pulsars 
\cite[e.g.][and references therein]{kulkarni88,naylor93,deloye08}, or 
ultracompact binaries \cite[e.g][]{nelemans01,belczynski02}.

Many papers that appeared later on were devoted to the population 
synthesis of LMXBs \cite[see e.g.][]{belczynski02,pfahl03,
vandersluys05,fragos08}, which until now suffer from uncertainties in 
the binary star evolution 
(e.g. the treatment of the common envelope stage, parameters of 
magnetic braking, etc.). Clearly, such detailed studies are very 
important and must be continued as they may help to constrain values of 
specific parameters of binary evolution from comparison with
observations. However, to 
%
understand the observed gross properties of the LMXB LF we can try to
use very general assumptions about binary star evolution, as we show
below. 

Consider an ensemble of binaries that has been formed in a galactic
halo over a time period $T$ in the past. Let $\tau(P)$ be the duration 
of an active LMXB stage of a source within the orbital period range 
$P, P+dP$. Let $n(P)$ be the number distribution of sources within
this period range. Assuming the steady formation of sources over time 
period $T$, the luminosity distribution can be written in the form 
$$
{dN\over{dL_{\rm x}}}\propto \frac{\tau(P)}{T} n(P) {dP\over{dL_{\rm x}}}\,.
$$
(Note that the first factor $\tau/T$ would not appear for a LF 
constructed from a sample of sources produced in an instantaneous star
formation burst; however, this is the case neither for the galactic 
bulge nor for elliptical galaxies we consider here.)

Clearly, the shape of LF will be determined by the dependence of all 
three factors in this expression on the X-ray luminosity. Consider 
these factors separately for bright (with giant donors) and dim (with
main-sequence donors) LMXBs.

At luminosities below $L_{\rm x} \approx 10^{37}$ \lum\ ($\dot{M}<10^{-9}$
$M_\odot$/year), the majority of donor stars in our LMXB sample are
main sequence stars with long life times ($>$Gyr). The stable mass 
transfer episode is also very long, so the factor $\tau/T$ should not 
strongly depend on $L_{\rm x}$. The period distribution $n(P)$ is
generally determined by the
evolutionary history of binaries which includes as a minimum a
supernova explosion to produce the neutron star, a common envelope
phase, etc. (or involves dynamical processes in dense stellar
clusters), and can be found for example from population synthesis
studies. Let us put it in the form $dN/d\log P$ and leave it as it is 
for a while.

The luminosity of a LMXB due to magnetic stellar wind braking depends 
on the orbital period as a power law  $L_{\rm x}\propto P^{2.5}$ 
\citep[see e.g.][]{iben84} or $L_{\rm x}\propto P^{3.3}$ 
\citep{patterson84}, or $L_{\rm x}\propto P^{3.8}$ \citep{cote89} so
written in the form $d\log P/d\log L_x$ this factor is independent 
of $L_x$.  So we find
$$
n(L_x)\propto  {1\over{L_{\rm x}}} {dN\over{d\log P}}{d\log P \over{d\log L_{\rm x}}} \sim  {1\over{L_{\rm x}}} {dN\over{d\log P}}
$$

This would match the observed shape of the LF of dim LMXBs if  the 
factor $dN/d\log P\sim$~const($L_x$). Unfortunately, it is difficult 
to construct this distribution from observations due to many selection 
effects. Population synthesis calculations \cite[e.g.][]{fragos08} 
produce a variable distribution, which only roughly can be considered 
as constant. Observations suggest that initial orbital periods of
binary stars do follow this dependence, i.e. $dN/d\log P=$const 
\citep{popova82}, and we shall assume this to hold approximately at
all stages of binary evolution.

LMXBs with luminosities higher than $L_{\rm x}>10^{37.5}$ \lum\ have 
predominantly giants companions (see above). The duration $\tau$ of
the mass transfer (and hence the accretion stage)  in these systems
is significantly smaller than that for systems with main sequence
donors, and to a large extent is determined  by the mass transfer rate 
\cite[see e.g.][]{webbink83}. Let us assume that the luminosity of
LMXB is a powerlaw function of its orbital period 
$L_{\rm x}\propto P^{\alpha}$ (which is true for the magnetic stellar
wind braking dominated regime, $\alpha\sim2.5-3.8$, see above, and 
also holds when nuclear evolution of the giant donor is responsible 
for its Roche lobe overflow, $\alpha\sim1$, see \citealt{webbink83}). 
We also adopted that the duration of the bright LMXB stage is 
inversely proportional to its luminosity $\tau \propto L_{\rm  x}^{-1}$. 
This is approximately correct because the total mass, which might be 
accreted from a late-type giant donor stars does not vary much from 
system to system \cite[see e.g.][]{webbink83}. Combining these we obtain
$$
{dN\over{dL_{\rm x}}} \propto  {1\over{L_{\rm x}P}}{dP\over{dL_{\rm x}}}\propto {1\over{L_{\rm x}^2}} {d \log P\over{d \log L_{\rm x}}}\approx L_{\rm x}^{-2}\,.
$$

Thus {\sl almost independently of the mechanism which drives Roche 
  lobe overflow in luminous LMXBs (with late-type giants) we obtain a
  slope of the LF similar to the observed one} 
\citep{gilfanov04,kim04}. The condition for this to be correct is that
the donor stars in these LMXBs are short-living giants. We can try to
estimate the longest lifetimes of such giants assuming that the
maximum mass, which might be accreted from them is $\la 0.6 M_{\odot}$ 
\cite[e.g.][]{webbink83} and their duty cycle is close to unity. In 
this case 
$\tau_{\rm giants}\la 0.6M_\odot/(2\times10^{-9}M_\odot/$year$)\sim3\times10^{8}$ 
years. This means that the list of bright LMXBs in the Galaxy should 
alter after approximately 30 Myrs. 

We would like to note here that the boundary between different types
of donors at luminosities above and below $L_{\rm x}\sim10^{37}$ erg/sec 
was also previously noted in numerical simulations of LMXB populations 
in \cite{fragos08,kim09}, but the slopes of the LMXB LF above and
below this luminosity was found to be not so different. The inability 
of particular numerical simulations to reproduce the observed break in 
the X-ray luminosity function of the LMXB population might indicate 
that we do not (yet) understand all details of the physical processes 
(e.g. the mode of the angular momentum loss, the common envelope 
phase, etc.) that shape the LMXB LF. We plan to perform more detailed 
calculations of the LMXB population in our future work.

\textit{Acknowledgments}. This work is partially supported by RFBR 
grants 10-02-00492-a, 10-02-00599-a, by Russian Federation President 
programm NSh-5069.2010.2, Federal Russian Science Agency through the 
research contract 02.740.11.0575, and program of Presidium RAS P-19  
and OFN-16.

\end{document}